\documentclass[showpacs]{revtex4}
\usepackage{graphicx}

\begin {document}

\title {A study on the turbulent transport of an advective nature in the fluid plasma}

\author{Byung-Hoon Min, Chan-Yong An and Chang-Bae Kim}
\email{cbkim@ssu.ac.kr}
\affiliation{Physics Department,
  Soongsil University, Seoul, KOREA} 

\date{\today}

\begin {abstract}
\indent 
Advective nature of the electrostatic turbulent flux of plasma energy is studied numerically in a nearly adiabatic state. Such a state is represented by the Hasegawa-Mima equation that is driven by a noise that may model the destabilization due to the phase mismatch of the plasma density and the electric potential. The noise is assumed to be Gaussian and not to be invariant under reflection along a direction $\hat s$. It is found that the flux density induced by such noise
is anisotropic: While it is random along $\hat s$, it is not
along the perpendicular direction ${\hat s}_\perp$ and the flux is not diffusive.  The renormalized response may be approximated as advective with the velocity being proportional to $(k\rho_s)^2$ in the Fourier
space $\vec k$.  

\end {abstract}



\maketitle

\section {Introduction}

Hasegawa-Mima (HM) model of the electrostatic fluctuations in the turbulent plasmas is a minimal foundation for the dynamics of the nonlinear interactions of drift waves in the adiabatic state \cite{HM}. The emergence and the subsequent role of the zonal flow is just one of many areas of the complicated plasma transport where the HM dynamics successfully provide pedagogical explanations \cite{PHD98,KK00}. For large scale fluctuations, however, the plasma is not believed to be adiabatic. The so-called Terry-Horton (TH) equation supplements the HM model by adding ad hoc non-adiabatic electron response as a way of making the perturbations linearly unstable due to inverse Landau damping \cite{TH82}. With such linear growth and artificial damping one may study the saturated states resulting from the dual cascades of the energy and the enstrophy. The HM model may be extended to include the collisional aspect of the drift waves by not neglecting the electron motion along the magnetic field line so that linearly unstable modes are present self-consistently and that the energy is transferred to locally adjacent scale and is eventually dissipated at small scale \cite{HW83,Terry00}. Such collisional drift wave (CDW) description of the plasma may be shown to reduce to the HM equation in the adiabatic limit. For a quantitative analysis of the CDW model numerical simulation is necessary. However, the fluctuations of the CDW model that are close to the adiabatic state are very slow to reach a saturation numerically. 

Instead of dealing with the self-consistent CDW model in assessing the non-adiabatic effect on the transport one may choose to approach conceptually similarly to the TH equation but more realistically by introducing a simulated noise that represents energy pumping of the linearly unstable CDW's. As CDW's are not isotropic, such noise needs to reflect the dependency on the directions. 
Specifically we assume that the noise is Gaussian with the spectrum 
of power-law type that is widely used in the renormalization-group
(RG) approach to turbulence \cite{AdzhemyanBook, CBK10}. 
The noise is assumed to be short-correlated in time
compared with typical time scale of the plasma fluctuations as well as
reflection non-invariant or parity non-conserving (PNC) along a
direction $\hat s$ as a symmetry breaking element. It has been
shown through RG calculations that such PNC noise leads to the
renormalized plasma response of advective nature.
This work is, thus, an extension of Ref.~\cite{CBK10} by
including numerical aspect to the analysis.
The present work finds out that the induced energy flux can be, indeed,
advective. The direction, however, is not along $\hat s$, but along
the perpendicular direction to $\hat s$. In Sec.~\ref{simulation} the model and
the results of the numerical simulation are described and we conclude in Sec.~\ref{conclusion}.

\section {Simulation model and results}
\label{simulation}

The forced HME for the electrostatic field $\varphi$ in the Fourier space is
 \begin {equation}
\partial _t \left( 1  + k^2 \right)\varphi _k  +
i k_y \varphi _k  - {\textstyle{1 \over 2}}\sum\limits_{\vec k =
  \vec p + \vec q } { M_{\vec k,\vec p,\vec q}
\varphi _p \varphi
  _q }  + \nu  k^4 \varphi _k  = f_k, 
 \label {HME}
 \end {equation}
%
%
%
%
%
%
where the space scale is normalized to $\rho_s$ and the time unit is
$\rho_s / v_d$ with the electron diamagnetic drift speed $v_d$,   
$M_{\vec k,\vec p,\vec q}  = \left( {\hat z \cdot \vec p \times \vec
    q} \right)\left( {p^2  - q^2 } \right)$ is the coupling between
three modes, and $\nu$ is the viscosity. The Gaussian random
forcing $f_k (t)$ is delta-correlated in time and the power
spectrum~$P$ of the noise in the $d$-dimensional Fourier space 
\begin{equation}
P(\vec k)= D k^{6-d-2\delta}\left( 1+ i \alpha\hat s\cdot \hat k \right).
\label{power spectrum}
\end{equation}
As the sum of $P$ over all $\vec k$ is proportional to the input rate
of the energy, $W=(1/2)\sum_{\vec k} \left( 1 + k^2 
\right) \left|\varphi_k\right|^2$, the power-law form of $P$, with a
proper exponent $\delta$, may be thought of as an approximation to a
part of localized realistic power spectrum. The coefficient $\alpha$
in Eq.~(\ref{power spectrum}) measure the strength of the PNC noise.

\begin{figure}
\begin{center}
\includegraphics[width=4in,keepaspectratio,clip, angle=-90]{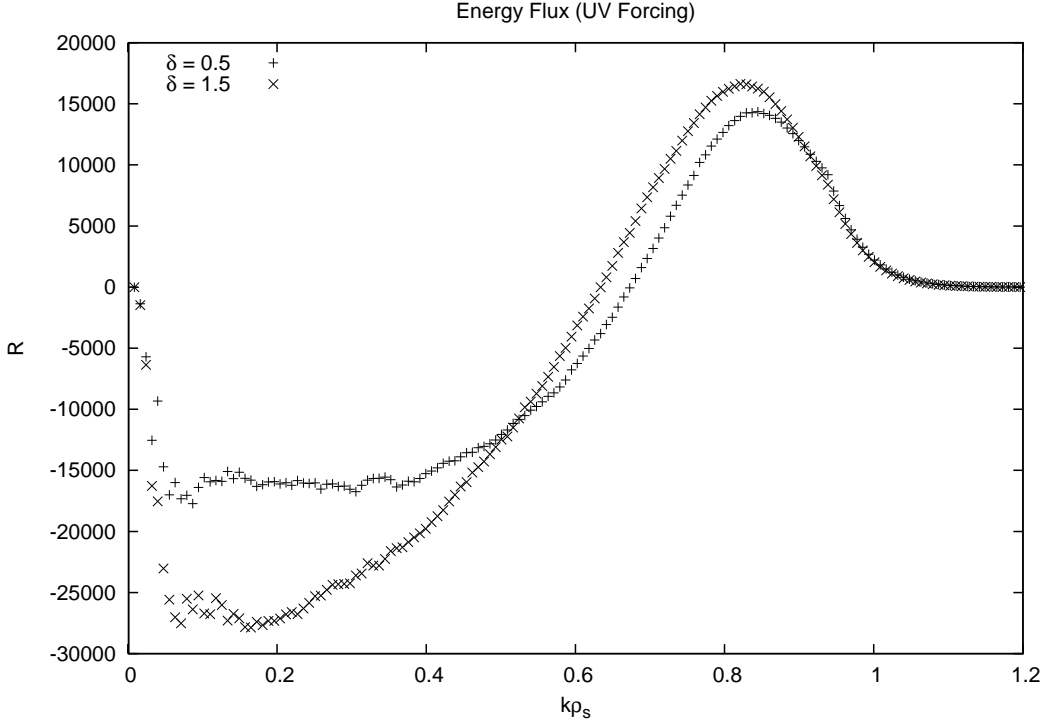}
\caption{Real parts of typical turbulent energy flux $R(k)$ with $\alpha=0.5$ is shown for two cases of uv pumping, $\delta=0.5$ and $\delta=1.5$ with the PNC direction along $\hat y$.}
\label{fig1}
\end{center}
\end{figure}

Eq. (\ref{HME}) is numerically integrated in time by advancing
a time step $\Delta t$ with Runge-Kutta 2nd-order scheme starting from
the zero initial conditions.   
If $\Delta t$ is taken to be shorter than the auto-correlation time $\tau$ of the
noise, the $\delta(t-t')$ part of the auto-correlation function of the
noise is approximated as $1/\Delta t$. The noise is numerically
generated by using two variables $\theta$ and $\xi$ \cite{Alvelius, Boffetta02}. 
They are uniformly distributed over $(0,2\pi)$ such that 
\begin{equation}
\left< \theta_{\vec k}(t) \theta_{\vec k'}(t+
\Delta t)\right> = \left<
  \xi_{\vec k}(t) \xi_{\vec k'}(t+\Delta t)\right> =  0 
\end{equation}
and between different $\vec k$'s they are independent of each
other. However, $\theta_{\vec k}(t)$ and $\xi_{\vec k}(t+\Delta t)$
are assumed to be correlated according to Eq.~(\ref{power spectrum}).  
The noise is, then, expressed as 
\begin{equation}
f_{\vec k} (t)= a_{\vec k} e^{i\theta_{\vec k}(t)} + i b(\vec k) e^{i\xi_{\vec k}(t)}
\end{equation}
and between two successive time steps the auto-correlation becomes
\begin{eqnarray}
\left< f_{\vec k} (t) f_{\vec k'}^* (t+\Delta t)
\right> & =& \left \{
|a_{\vec k}|^2 + |b_{\vec k}|^2 - i a_{\vec k} b_{\vec k}^* \left< e^{i[\theta_{\vec
      k}(t)-\xi_{\vec k}(t+\Delta t)]}  \right> \right.
\nonumber
\\*
&&\quad \left.
+ i a_{\vec k}^* b_{\vec k} \left< e^{i[\xi_{\vec k}(t)-\theta_{\vec
      k}(t+\Delta t)]}  \right> \right\}\delta_{\vec k \vec k'}
\label{ff}
\end{eqnarray}
It is convenient to choose $\theta_{\vec k}(t) = \xi_{\vec k}(t+\Delta
t)$.  Then, $\left< e^{i[\theta_{\vec k}(t)-\xi_{\vec k}(t+\Delta t)]}
\right>=1$ and $\left< e^{i[\xi_{\vec k}(t)-\theta_{\vec k}(t+\Delta
    t)]}  \right>=0$. Thus, from Eqs. (\ref{power spectrum}) and
(\ref{ff}), $a_{\vec k}$ and $b_{\vec k}$ are
determined by 
\begin{equation}
a_{\vec k}^2 + b_{\vec k}^2 - i a_{\vec k} b_{\vec k} = D 
k^{6-d-\delta}(\Delta t)^{-1} (1+i \alpha \hat k\cdot\hat s).
\end{equation}
\begin{figure}
\begin{center}
\includegraphics[width=4in,keepaspectratio,clip, angle=-90]{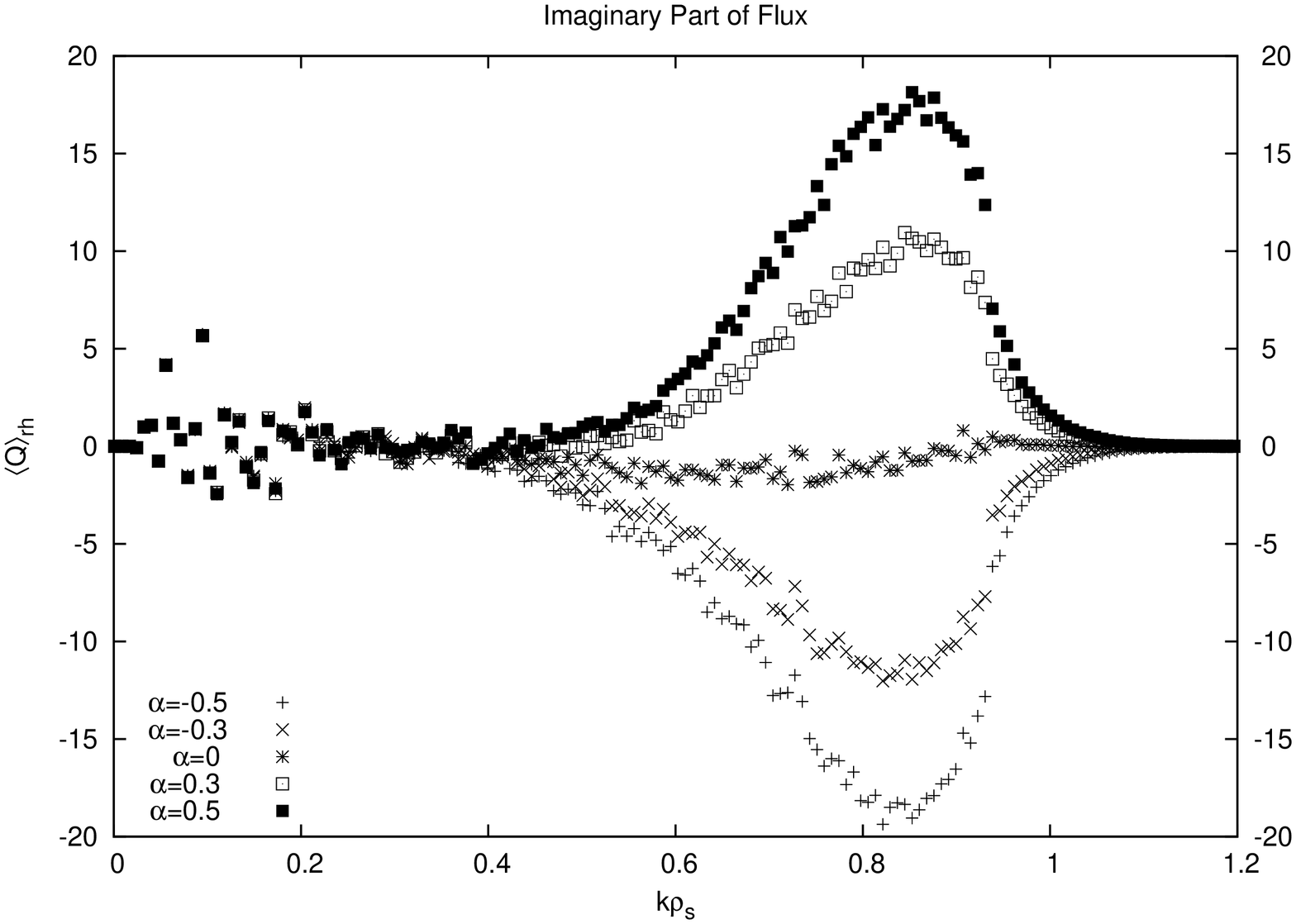}
\caption{Imaginary parts of the enstrophy turbulent flux $\left< Q
  \right>_{\rm rh}$ in the right half plane $k_x >0$ are plotted
  versus $k$ for the cases of $\alpha=0, \pm0.3, \pm 0.5$ with
  $\delta=0.5$ and $\hat s = \hat y$.} 
\label{fig2}
\end{center}
\end{figure}

A case of uv forcing is simulated where the noise is applied in
the scale that is smaller than $\rho_s^{-1}$, that is $k_f  < 1$ in our unit of
$k$ being equal to $\rho_s^{-1}$. We set  $L=2\pi\times1024$, $N=4096$
and consider a uv forcing of $\delta=0.5$ with amplitude
$f_0=\sqrt{D}=0.5$ in the range $1/1024 \le k \le 960/1024$ with $\hat
s = \hat y$. Additional parameters are  
the time step $\Delta t = 5\times 10^{-4}$ and the
viscosity $\nu=10^{-4}$. Hyper-viscose damping term of the form 
$4.4\times10^{2} k^{16}\varphi_k$ is introduced in Eq.~(\ref{HME}) that localizes uv forcing. The
Eckman-like friction term $0.005 \varphi_k$ is added in Eq.~(\ref{HME})
that, by providing a necessary damping, secures the saturation of the
energy transfer toward smaller $k$ \cite{Mazzino}. 

Figs. \ref{fig1} and \ref{fig2} are collections of plots regarding the
turbulent flux. In Fig. \ref{fig1}, the typical energy flux expressed in
terms of $k=| \vec k |$,   
$R(k) = {\textstyle{1 \over 2}}\sum\limits_{k' = 0}^{k} \sum\limits_{\vec k' =
  \vec p + \vec q } { M_{\vec k' ,\vec p,\vec q}
\varphi_{-k'} \varphi_p \varphi_q }$  with $\delta=0.5$ and $\alpha=0.5$
is shown. 
The negative sign of $R$
means that the energy flows toward smaller $k$. Roughly in the region
$0.04\le k \le 0.4$ inertial range of constant energy flux is seen to
form. For a comparison, the flux of another uv forcing, $\delta=1.5$, is also plotted where the inertial range is not as vivid as the former case. It is checked that the shape of $R(k)$ is almost independent of the measure of the PNC noise $\alpha$ as expected. 
\begin{figure}
\begin{center}
\includegraphics[width=4in,keepaspectratio,clip, angle=-90]{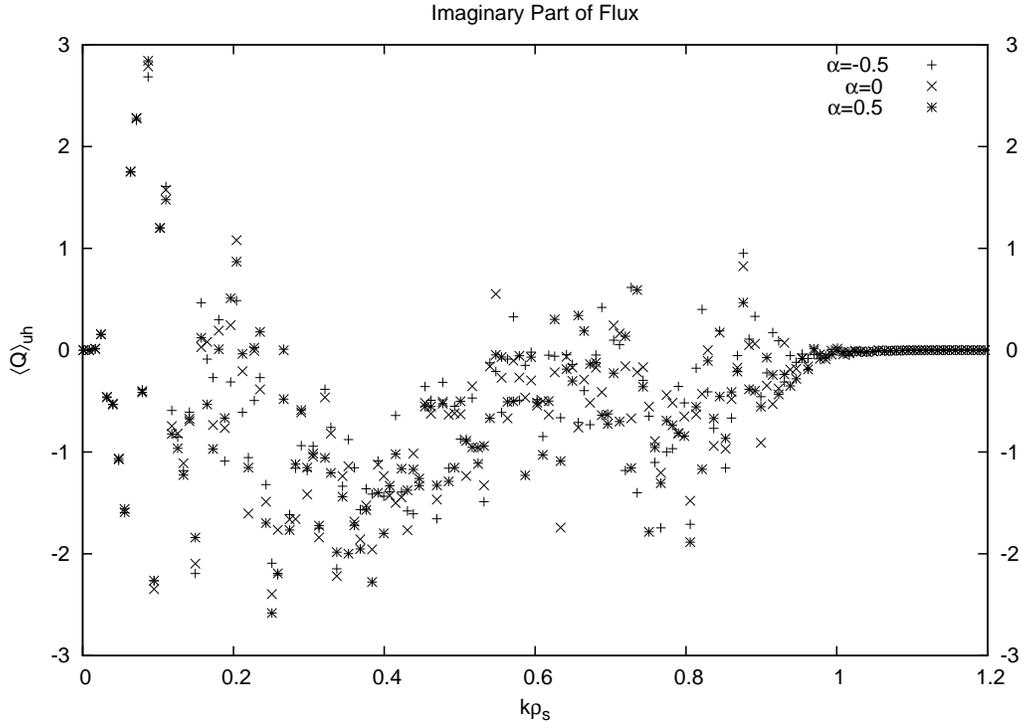}
\caption{Imaginary parts of the enstrophy turbulent flux $\left< Q
  \right>_{\rm uh}$ in the upper half plane $k_y >0$ are plotted
  versus $k$ for the cases of $\alpha=0$ and $\pm 0.5$ with
  $\delta=0.5$ and $\hat s = \hat y$. 
}
\label{fig3}
\end{center}
\end{figure}
The effects of the PNC noise $\alpha$ in the flux 
are observed in Fig.~\ref{fig2} where the imaginary part of the
turbulent enstrophy flux is plotted. The enstrophy flux density $Q$ is obtained upon multiplying both sides of Eq. (\ref{HME}) by $\omega_{-k}$. If it is summed over all $\vec k$'s, its imaginary part cancels out. Therefore, In order to elucidate the role of the imaginary part of the flux, partial sums are taken over a semi-annular disks of small width $\delta k$ and shown in Fig.~\ref{fig2}: $Q_{rh}(k)$ in the right-half plane, $k_x \ge 0$, and $Q_{uh}(k)$ in the upper-half
plane, $k_y \ge 0$. The flux $Q_{rh}(k)$ is defined as 
\begin{equation}
Q_{rh}(k) = \Im {\textstyle{1 \over 2}}\sum\limits_{k'
  = (k-\delta k)}^{k} \sum\limits_{k'_x\ge0} { M_{\vec k' ,\vec p,\vec q} {k'}^2
\varphi_{-k'} \varphi_p \varphi_q }
\label{Q_rh}
\end{equation}
and $Q_{uh}(k)$ correspondingly
by the substitution of $\sum\limits_{k'_x\ge0}$ with $\sum\limits_{k'_y\ge0}$. It is implied in Eq. (\ref{Q_rh}) that $\vec k' =  \vec p + \vec q$ in the sum. Both $Q_{rh}(k)$ and $Q_{uh}(k)$ are the enstrophy fluxes over the corresponding half annulus of width $\delta k$, which is 1/1024 in this case, in the $(k_x, k_y)$ plane.
Shown in Fig.~\ref{fig2} are the mean values of $\left< Q \right>_{\rm rh}$ in terms of $k$ for the cases of $\alpha
= 0, \pm0.3 $ and $\pm 0.5$ being averaged over 200 sampled data that are
taken at every $100\Delta t$. 
Since for the isotropic
forcing ($\alpha=0$) $\left< Q \right>_{\rm rh}$ is negligibly zero,
$\left< Q \right>_{\rm rh}$ with $\alpha\neq0$ is purely induced by the
PNC noise.
The size of $\left< Q \right>_{\rm rh}$
is proportional to $\alpha$ and it changes sign correspondingly to the
sign of $\alpha$.

The fluxes in the upper-half plane, $k_y\ge0$, $\left< Q \right>_{\rm uh}$ are plotted in
Fig. \ref{fig3} for $\alpha=0$ and $\pm 0.5$. Upon comparison with Fig. \ref{fig2}, 
it is observed that,  for both
the isotropic ($\alpha=0$) and the PNC forcing irrespective of
the sign of $\alpha$, $\left< Q \right>_{\rm uh}$ seems to be little
different from one another. It is clear that such anisotropic nature of $\left< Q
\right>_{\rm uh}$ is the result of the nonlinear interactions between
the drift waves and the PNC noise is not relevant to $\left< Q
\right>_{\rm uh}$ \cite{Nazarenko09}.  Since the Hasegawa-Mima dynamics has an intrinsic direction of $\hat y$
due to the gradient of the plasma density along $\hat x$, it is
understandable for isotropic forcing that $Q$ has odd parity under the
reflection along $\hat y$ leading  $\left< Q \right>_{\rm rh}$ to
vanish while leaving $\left< Q \right>_{\rm uh}$ finite. It may be
concluded that the turbulent flux induced by the PNC noise is
predominantly along the perpendicular direction to $\hat s$, that is
$\hat x$ in the present setup.  

\begin{figure}
\begin{center}
\includegraphics[width=4in,keepaspectratio,clip, angle=-90]{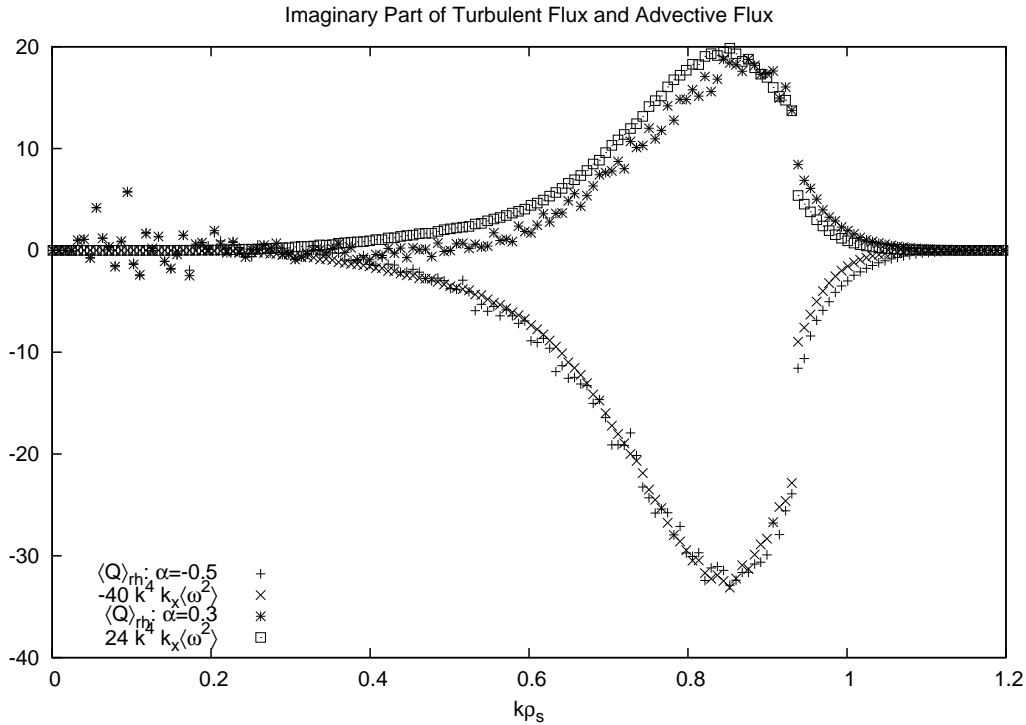}
\caption{Flux density $\left< Q \right>_{\rm rh}$ in the right-half
  plane is compared with the advective flux density $U \left<k_x
    \omega^2 \right>_{\rm rh}$ 
  where $U$ is advecting velocity along $\hat x$. $U$ is found to be proportional to $k^2$.}
\label{fig4}
\end{center}
\end{figure}

Since $\left< Q \right>$ in the left-hand plane $\left< Q \right>_{\rm
  lh}$ is equal to $-\left< Q \right>_{\rm rh} $, $\left<
  Q \right>$ suggests advective nature along the direction of $\hat x$,
perpendicular to the PNC direction. Upon anticipating $\left< Q
\right>_{\rm rh} $ to be advective with the velocity $\hat x U$, an
advective flux $ U \left< {k_x}\omega^2 \right>_{\rm rh} $, where
  $\omega_k$ is the vorticity $-k^2 \varphi_{k}$, in the right-half
plane is laid over $\left< Q \right>_{rh}$ in Fig.~\ref{fig4}. Scaling
  the 
advecting velocity $U$ with $k^2$ seems to fit $\left< Q
\right>_{\rm rh} $ to the advective flux. It is determined that, for $\alpha=-0.5$, $U=-40
  k^2$ and, for $\alpha=0.3$, $U=24 k^2$. One notes that $U$ is
  proportional to $\alpha$ and that the dependence of $U$ on $k^2$ agrees with
  the prediction of the RG and the strongly magnetized MHD plasmas.  

\begin{figure}
\begin{center}
\includegraphics[width=4in,keepaspectratio,clip, angle=-90]{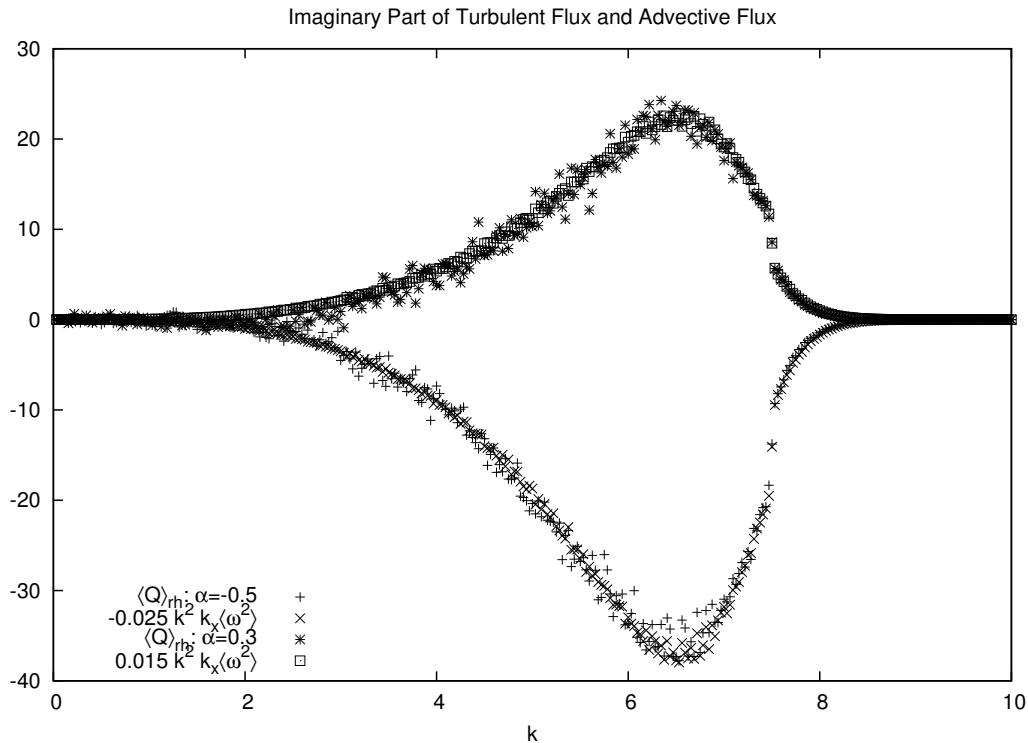}
\caption{PNC-induced turbulent enstrophy fluxes $\left< Q \right>_{\rm
    rh}$ for the cases of $\alpha=-0.5$ and 0.3 
  are plotted at $\delta=0.5$ and the box size $L=2\pi\times32 $. For
  comparison, advective fluxes $U k_x \left< |\omega_k|^2 \right>$ with velocity $U=-0.025 k^2$ and $U=0.015 k^2$ are overlaid correspondingly } 
\label{fig5}
\end{center}
\end{figure}

In order to check the validity of the $k^2$ dependence of the advecting veloctiy another case of uv forcing is considered where $k_f>1$. The box size
is $L=2\pi\times 32$, $N=1024$, the hyper-viscosity is set to be
$10^{-12}$ with the uv forcing of $\delta=0.5$ in the region
$1/32 \le k \le 240/32$ and $f_0 = 3.6\times 10^{-2}$. Similar to the former case of $k_f<1$ an inertial range is found to exist and the dependence of $\left< Q \right>$'s on $\alpha$ is verified. In 
Fig. \ref{fig5} $\left< Q \right>_{\rm rh}$  for 
the cases of $\alpha=-0.5$ and 0.3 are overlaid with $U k_x \left< |\omega_k|^2 \right>$ where the 
advecting velocity is $U=-0.025 k^2$ and $U=0.015 k^2$, respectively.  
One may, thus, conclude that an effect of the PNC noise (\ref{power spectrum}) on the convective nonlinear term in Eq. (\ref{HME}) may result in the advective response. With the PNC direction along $y$, regarding the transport of the second-order moment like the energy and the enstrophy, one may approximate Eq. (\ref{HME}) as
 \begin {equation}
\partial _t \left( 1  + k^2 \right)\varphi _k  +
i \left( k_y  +  U k_x k^2 \right)\varphi_k  + \nu  k^4 \varphi _k  \approx f_{\rm PC}, 
 \label {HME renorm}
 \end {equation}
where $U\propto k^2$ and $f_{\rm PC}$ is the parity-conserving part of the noise.

\section{Conclusion}
\label{conclusion}

Motivated by observations that the fluctuations in the magnetic fusion devices are predominantly anisotropic as attested by the presence of various types of the drift waves, we set out to pedagogically study the advective characteristics of the turbulent transport. In this paper we 
simulate the dynamics of the
electrostatic fluctuations in a magnetized plasma modeled by
Hasegawa-Mima equation that is driven by an anisotropic noise. The noise is not
reflection-invariant along the direction $\hat s$. Such noise is
simulated numerically by considering, for each
mode, two random variables that have time-delayed correlation to each
other for a short time in a time-ordered fashion. 

To be specific the
electrostatic fluctuations are considered to be driven by the noise in
the uv scale so that the plasma energy cascades toward smaller scale. 
In the region of
the Fourier space $\vec k$ where the energy flux is nearly constant
owing to the uv noise power it is found that the imaginary part of the
turbulent flux $\left< Q \right> $ is advective.  With the case of $\hat s = \hat y$, the advective flux $\left< Q \right>_{\rm uh} $ along the direction $\hat y$ is arguably due to the strong interactions of the drift waves based on the fact that $\left< Q \right>_{\rm uh} $ is mostly independent of the PNC noise. However, the advective flux $\left< Q \right>_{\rm rh} $ along the direction $\hat x$ arises because of the PNC noise. The advecting velocity $U$ along $\hat x$ is found to be proportional both to $(k\rho_s)^2$ and
to the strength of the parity-non-conserving noise. Such dependence is shown to
hold in a wide range of $k$, including $k\rho_s<1$, as long as the
plasma is forced from a uv region. 
The $k^2$ dependence of $U$ is in accordance with the strongly magnetized MHD plasma forced by a PNC noise. As it is clear that the PNC noise may lead to the non-diffusive transport in the plasma, it is desirable to look into the presence of PNC properties in the turbulent plasmas. Work in such a direction is under way and results will be presented in another publication. 


%
\begin{acknowledgments}

This work was supported by the Korea Research Foundation Grant funded
by the Korean Government (2009-0082669).  

\end{acknowledgments}


\section*{References}

\begin {thebibliography}{10}

\bibitem{HM}
A. Hasegawa and K. Mima, Phys. Fluids {\bf12} 87 (1978).

\bibitem{PHD98}
P. H. Diamond, M. N. Rosenbluth, F. L. Hinton, M. Malkov, J.
Fleischer, and A. Smolyakov, International Atomic Energy
Agency, Vienna, Report No. IAEA-CN-69/TH3/1, 1998 (unpublished)

\bibitem{KK00}
J. A. Krommes and C.-B. Kim, {Phys. Rev. E} {\bf62} 8508 (2000).

\bibitem{TH82}
P. Terry and W. Horton, Phys. Fluids {\bf25} 491 (1982).

\bibitem{HW83}
A. HWaegawa and M. Wakatani, Phys. Rev. Lett. {\bf50} 682 (1983).


\bibitem{Terry00}
P. W. Terry, {Rev. Mod. Phys.} {\bf 72} 109 (2000).

\bibitem{AdzhemyanBook}
L. Ts. Adzhemyan, {\it The Field Theoretic Renormalization Group in
 Fully Developed Turbulence} (Gordon and Breach Sci. Pub., Amsterdam, 1999).

\bibitem{CBK10}
C. B. Kim,  {Nucl. Fusion} {\bf50} 045001 (2010).

\bibitem{Alvelius}
K. Alvelius,  {Phys. Fluids} {\bf 11} 1880 (1999).

\bibitem {Boffetta02}
G. Boffeta, A. Celani, S. Musacchio and M. Vergassola {Phys. Rev.} E {\bf 66} 026304 (2002).

\bibitem {Mazzino}
A. Mazzino, P. Muratore-Ginanneschi and S. Musacchio, {Phys. Rev. Lett.} {\bf 99} 144502 (2007); A. Mazzino, P. Muratore-Ginanneschi and S. Musacchio ,  {J. Stat. Mech.} P10012 (2009).

\bibitem{Nazarenko09}
S. Nazarenko and B. Quinn  {Phys. Rev. Lett.} {\bf103} 118501 (2009).






















\end {thebibliography}
\end{document}